\documentclass[twocolumn,prd,nofootinbib,superscriptaddress]{revtex4}
\usepackage{color}
\usepackage{amsmath}
\usepackage{mathrsfs}
\usepackage{graphicx}
\usepackage{booktabs}
\usepackage{bm}
\usepackage{hyperref}

\newcommand\SkipNeuralNets[1]{}
\newcommand\skipme[1]{}

\newcommand\optional[1]{}
\newcommand\unit[1]{{\rm #1}}

\newcommand\mc{{{\cal M}_c}}

\def\RIT{Center for Computational Relativity and Gravitation, Rochester Institute of Technology, Rochester, New York
  14623, USA}
\def\Fullerton{Nicholas and Lee Begovich Center for Gravitational-Wave Physics and Astronomy,
California State University Fullerton, Fullerton, California 92831, USA}

\begin{document}

\title{Inference on neutron star parameters and the nuclear equation of state with RIFT, using  prior EOS information }
\author{A. Vilkha}\affiliation {\RIT}
\author{A. Yelikar}\affiliation {\RIT}
\author{R. O'Shaughnessy} \affiliation{\RIT}
\author{J. Read} \affiliation{\Fullerton}

\begin{abstract}

In this paper, we present an inference method for determining neutron star parameters and constraining the nuclear equation of state (EOS) using the RIFT parameter inference engine. 
We incorporate externally-produced prior information about the EOS to improve the accuracy and efficiency of the inference process. 
We apply this method to the GW170817 event and assess its performance. 
Our results demonstrate the effectiveness of incorporating prior EOS information in the inference process, leading to sharper conclusions and more rapid inference on new detections. 
This approach has the potential to enhance our understanding of neutron stars and the nuclear EOS in future gravitational wave observations.

\end{abstract}

\maketitle

\section{Introduction}
\label{sec:intro}

The nuclear equation of state, which  characterizes the relationship between pressure and density of nuclear matter, has
been recently increasingly well constrained by multiple theoretical and observational techniques in the last few years.
Techniques from  chiral effective field theory, from state-of-the-art crust models, and from perturbative-QCD
limits \cite{2010PhRvD..81j5021K, 2020NatPh..16..907A} have provided new EOS constraints and modeling options at low and
high density.  
Measurements of galactic neutron star masses and radii from  NICER and other radio and X-ray measurements provide
critical complementary perspectives, bounding other features of the EoS; see, e.g., \cite{2021ARNPS..71..433L,2022NatRP...4..237Y,gwastro-ns-eos-Kedia2024,2021ApJ...922...14P}  and
references therein.
Most famously, the BNS merger GW170817/AT2017gfo was indentified as a gravitational wave signal by the Advanced LIGO \cite{TheLIGOScientific:2014jea} and Advanced Virgo \cite{TheVirgo:2014hva} detectors,
as well as electromagneticaly by multiple telescopes \cite{LIGO-GW170817-mma,LIGO-GW170817-bns,LIGO-GW170817-EOS,LIGO-GW170817-EOSrank}, 
providing suggestive insight into the highest nuclear densities
\cite{LIGO-GW170817-bns,LIGO-GW170817-SourceProperties,LIGO-GW170817-EOS, 2020GReGr..52..109C, Landry:2020vaw,2023PhRvD.107d3034M,%
Reed:2021nqk,Essick:2021kjb,%
Agathos:2015uaa, Lackey:2014fwa, Alvarez-Castillo:2016oln, Margalit:2017dij, Annala:2017llu, Most:2018hfd,%
Radice:2017lry, Rezzolla:2017aly, Riley:2018ekf, Tews:2018iwm, Greif:2018njt, Kiuchi:2019lls, %
Landry:2018prd, Shibata:2019ctb, 2020NatAs...4..625C, Pang:2020ilf, 2020ApJ...893...61Z, Essick:2020flb,%
2020ApJ...893L..21R, Dietrich:2020efo, 2020EPJST.229.3663C, 2020PhRvC.102d5807L, 2021PhRvC.103c5802X,%
2021PhRvL.126f1101A,2019AnPhy.41167992H}. 
While customarily the inital GW from merging NS are interpreted in a highly model agnostic fashion, in isolation from
any other prior insight, the conclusions available from any individual future observation are the
sharpest when informed by strong physical priors.   
For this reason, several groups have already presented results on GW170817 using these external insights as  prior
information about the nuclear equation of state; see, e.g.,
\cite{2020NatAs...4..625C,2020Sci...370.1450D,2021PhRvD.104f3003L,2024PhRvC.109b5807P,2024arXiv240204172K,LIGOScientific:2019eut,Huth2022,Ray2023}  and
references therein.
In particular, these calculations all rely on large sets of many full EOS realizations.
Operationally and algorithmically speaking, these tabular approaches allow for direct input from a wide variety of
heterogeneous techniques to generate plausible EOS realizations and apply various theory-motivated constraints.
Though effective, when examined in detail these calculations have some weaknesses in efficiency and flexibility. 
Specifically, many approaches rely on comparison of each EOS to a fiducial,
benchmark inference of a candidate event (here, some reference analysis for GW170817), using density estimates
calibrated to their small set of output samples.  These inference methods can break down if too few fiducial samples are
consistent with our other assumptions.    Conversely, if the tabular EOS is simultaneously sampled during inference, due
to the high overhead associated with the gravitational wave likelihood for complex time-domain waveforms,  tabular EOS
inference may be limited to the simplest models for compact binary gravitational waves.  However, previous studies have
demonstrated notable BNS waveform systematics, at a level that can impair EOS inference in future and even present instruments \cite{2021PhRvD.103l4015G,2022PhRvD.105f1301K,2024arXiv240416599Y}.
Finally, many of the calculations have adopted various technical simplifications (e.g., fixing the
chirp mass; marginalizing over spins with an a priori reference spin distribution) which endanger their long-term
utility for  large-scale EOS inference from a NS population.

Recently, Wofford et al \cite{gwastro-RIFT-Update} described an extension of the RIFT parameter inference engine 
designed to incorporate externally-produced prior information about the nuclear equation of state, expressed in tabular form.
In this work, we assess this proposed method, as applied to GW170817.  
We demonstrate this approach can provide efficient prior-informed inference on binary properties, with substantial
impact on the nuclear equation of state.
This paper is organized as follows.  In Section \ref{sec:methods} we review the parameter inference methods, waveform
models, and strategy for EOS inference.  In Section \ref{sec:results} we demonstrate our method replicates previous
analysis of GW170817 using the same input EOS family.  We also apply our EOS inference technique to corroborate the
surprising result from Kedia et al \cite{gwastro-ns-eos-Kedia2024}: using different plausible astrophysical priors can produce notably different conclusions about
the EOS favored by GW170817.  We conclude in Section \ref{sec:conclusions} with brief comments on the impact of our investigation.

\section{Methods}
\label{sec:methods}

\subsection{GW inference}
\label{subsec:gw-inference}

A coalescing compact binary in a quasicircular orbit is described by its intrinsic
and extrinsic parameters.  
By extrinsic parameters, we refer to the seven numbers needed to characterize its spacetime location and orientation.
In practice, we use these  parameters only to determine how our instruments respond to a comprehensively-understood
single source; for example, the source orientation and distance influence can determine its apparent brightness at Earth.
By intrinsic parameters, we refer to the binary's  masses $m_i$, spins $\bm{\chi}$, and any quantities
characterizing matter in the system such as the dimensionless tidal deformabilities $\Lambda_i$.  These parameters are also the only physics needed to determine the nature of any emission from the system.
We adopt the conventions as previous studies to characterize the binary's parameters (e.g., coordinates, reference
frequencies, et cetera) \cite{gwastro-PE-AlternativeArchitectures,gwastro-PENR-RIFT}.
We will use ${\bm \lambda},{\bm \theta}$ to
refer to intrinsic and extrinsic parameters, respectively.
For simplicity, we will assume that either the source redshift is known or that it is small, so the detector-frame and
source-frame component masses are effectively equal (i.e., 
the differences in radii and tidal deformabilities due to redshift are small compared to both these values deduced from $m_{z,i}\simeq m_{}$ and the posterior width).

The RIFT gravitational wave inference strategy splits calculations that require
detailed source physics from fast, largely geometrical calculations needed to assess detector response.  This algorithm  consists of a two-stage iterative process to interpret gravitational wave observations $d$ via comparison to
predicted gravitational wave signals $h(\bm{\lambda}, \bm\theta)$.   In one stage, denoted ILE,
RIFT computes a marginal likelihood 
\begin{equation}
 {\cal L}\equiv\int  {\cal L}_{\rm full}(\bm{\lambda} ,\bm\theta )p(\bm\theta )d\bm\theta
\end{equation}
from the likelihood ${\cal L}_{\rm full}(\bm{\lambda} ,\theta ) $ of the gravitational wave signal in the multi-detector network,
accounting for detector response; see the RIFT paper for a more detailed specification.  This marginalized likelihood
depends only on source parameters, observational data, and the adopted waveform model relating the expected gravitational wave
signal observed at the earth to the source parameters.  For simplicity and technical convenience, in this work, we assume a well-identified electromagnetic
counterpart, with known source distance and sky location.
Likewise, for convenience and to be consistent with previous studies of GW170817, we adopt the \texttt{IMRPhenomPv2\_NRTidal}
waveform approximation \cite{2019PhRvD..99b4029D,2019PhRvD.100d4003D}, along with the assumption that the binary spins
are parallel to the orbital angular momentum.
Within this context, RIFT's  marginal likelihood calculation takes as input only the parameters of this waveform model
(the masses, spins, and  assumed tidal deformabilities $\Lambda_i$ of each object), and
is otherwise agnostic to the nuclear equation of state.

In the second stage, denoted CIP, RIFT first generates an approximation to ${\cal L}(\lambda)$ based on its
accumulated archived knowledge of marginal likelihood evaluations 
$(\lambda_\alpha,{\cal L}_\alpha)$.  Then, using this approximation, it deduces the (detector-frame) posterior distribution
\begin{equation}
\label{eq:post}
p_{\rm post}=\frac{{\cal L}(\bm{\lambda} )p(\bm{\lambda})}{\int d\bm{\lambda} {\cal L}(\bm{\lambda} ) p(\bm{\lambda} )}.
\end{equation}
where prior $p(\bm{\lambda})$ is the prior on intrinsic parameters like mass and spin.  
Since the conceivable nuclear equation of state imposes a relationship $\Lambda(m)$, this stage of RIFT's analysis must be
aware of the nuclear equation of state to perform self-consistent, EOS-informed analysis.

\subsection{Inference on the nuclear equation of state }
\label{subsec:eos-inference}

Following \cite{gwastro-RIFT-Update}, we attempt to infer the nuclear equation of state from a sequence
$\alpha=\left\{1\ldots N\right\}$ of equations of state by using a local order statistic: the fiducial tidal
deformability of a NS with equal masses and comparable detector-frame chirp mass: $S=\Lambda(\mc_z 2^{1/5})$ (though we
use the source-frame chirp mass if the host redshift is known).  We
perform inference on the NS properties using binary masses, spins, and this ordering statistic as our parameters,
constructing a posterior on $X=(m_i,\boldsymbol{\chi}_i,S)$.  At each stage of our analysis,  an approximate posterior on the nuclear
equation of state follows by postprocessing, identifying each $S$ with the closest element $S_\alpha$ in our collection
of equations of state. 

Using this approach, RIFT can  employ any EOS tabulation conditioned on any previous measurements within its usual
inference technique.  RIFT's analyses will be more efficient, focused on the (often narrow) range of tidal
deformabilities favored by contemporary families of EOS.  Moreover, because RIFT's output still saves all marginal
likelihood evaluations, our analysis can be used as (part of) any subsequent \emph{prior-uninformed} investigations over
a wider prior range.

Our approach makes two modest, well-motivated simplifying  approximations, allowing us to condense the effect of the EOS
into a single degree of freedom.  First, we assume that 
only  the
leading-order tidal deformability $\tilde{\Lambda}$ fully characterizes the effect of the EOS on emitted gravitational
waves, via our fit of the likelihood versus binary and tidal parameters.    Second, we effectively assume that the tidal
deformability $\tilde{\Lambda}$ varies slowly with mass ratio, such that binaries with comparable chirp mass have
comparable $\tilde{\Lambda}$ \cite{2018PhRvD..98f3020Z}. 
We test this assumption for validity in Appendix \ref{app:mass-ratio-dependence}.

\subsection{Prior information for the EOS}
\label{subsec:eos-prior}
\begin{figure}
\includegraphics[width=\columnwidth]{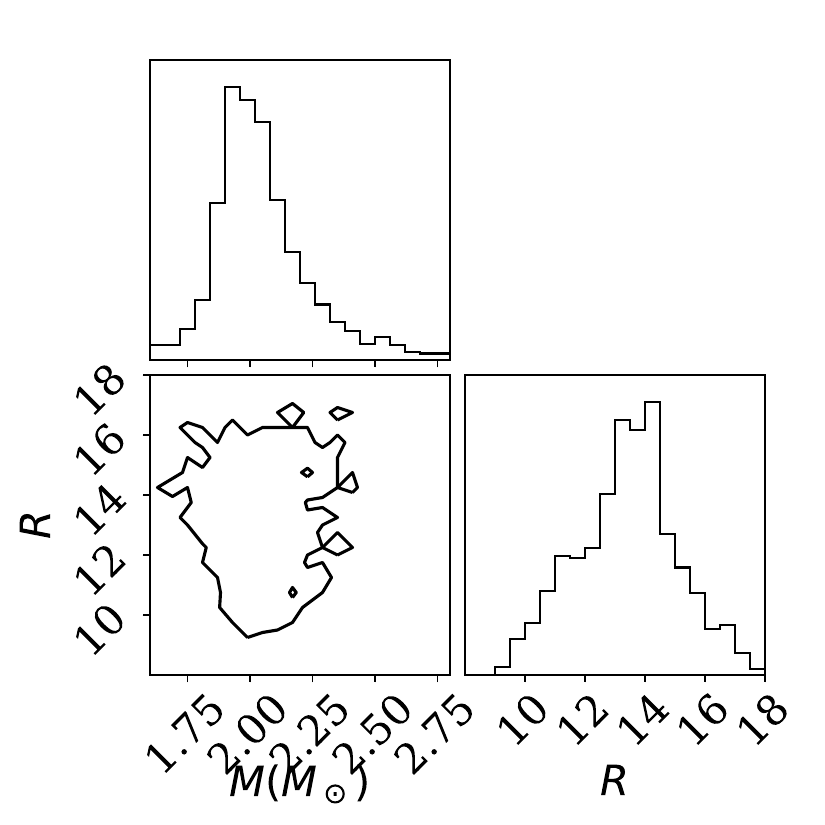}
\caption{\label{fig:eos_prior} \textbf{Prior on equation of state}: Quantities derived from the equation of state, for
  randomly selected equations of state drawn from the prior distribution provided by \cite{2021PhRvD.104f3003L} from
  pulsar observations.  Figure shows the joint distribution of the maximum neutron star mass $M_{\rm max}$ and radius $R_{1.4}$ of a
fiducial $1.4 M_\odot$ neutron star, given our prior assumptions.
}
\end{figure}

Several groups have provided extensive tables of nuclear equations of state, representing theoretical and
observational constraints arising from a variety of channels.  In this work, we adopt one fiducial analysis which
combines observations of pulsar masses, NICER mass-radius constraints, and GW observations \cite{2021PhRvD.104f3003L}.  This analysis
provides EOS instances both conditioned without and with GW observations. 
In this paper, we employ as our fiducial prior distribution an analysis conditioned on contemporary pulsar observations
from Legred et al
\cite{2021PhRvD.104f3003L}.   Their data product provides both EOS realizations relating pressure and density $p(\rho)$,
as well as  the inferred mass-radius trajectory $(M(\rho_c),R(\rho_c))$ deduced from the TOV equation for a range of
central densities.   
Figure \ref{fig:eos_prior} shows the distribution of selected neutron star quantities derived under these prior assumptions.

The raw Gaussian process EOS realizations provided by Legred et al allow for (and describe) multiple stable branches for
each EOS, with extrema in $M(\rho_c)$ that sometimes occur  both below $1M_\odot$ and regularly occur above $1.7 M_\odot$.
In this work, for each EOS we  restrict our investigation if needed to a single  branch containing neutron stars with
$M=1.4 M_\odot$. 
Our EOS inference effectively covers the mass range of $1.0-1.9 M_\odot$, which is consistent with many of the observed neutron star masses.  
Thus, if the extremum of the $M(\rho_c)$ curve is below $1.0 M_\odot$, we will not truncate that EOS realization 
since it does not influence the EOS inference in our analysis. 
However, if the extremum is present above $1.9 M_\odot$, we truncate the EOS realization at the extremum beyond the
density corresponding to $1.9
M_\odot$ to avoid  multiple  branches in the $M(\rho_c)$ curve. 
We also ensure that the single branch before the extremum is not truncated. 
While a small number of EOS models in our reference sample (about 20 out of $10^4$) have extrema in the range of
$1.0-1.9 M_\odot$, for simplicity we do not truncate these EOS as their inclusion has minimal impact on our results.

\subsection{Posterior for the EOS}
In our approach, the  posterior distribution produced by RIFT over $X$ implies a posterior distribution over
the conventional intrinsic parameters $m_i,\chi_i$ as well as the EOS indexing parameter $\alpha$.  In particular, each
posterior sample contains both the intrinsic parameters and an EOS index $\alpha$.  

Our inference procedure, if undersampled initially, may sparsely cover the EOS posterior.  To quantify how thoroughly we
cover the EOS space, we calculate an analog of $n_{\rm eff} $:
\begin{align}
n_{\rm eff, EOS} = N/\text{max}_\alpha \text{count}(\alpha)
\end{align}
where the denominator represents the largest number of times any individual EOS appears in our posterior sample.  

\section{Results}
\label{sec:results}

\subsection{Prototype analysis of GW170817}
\label{subsec:170817}
\begin{figure}
\includegraphics[width=0.9\columnwidth]{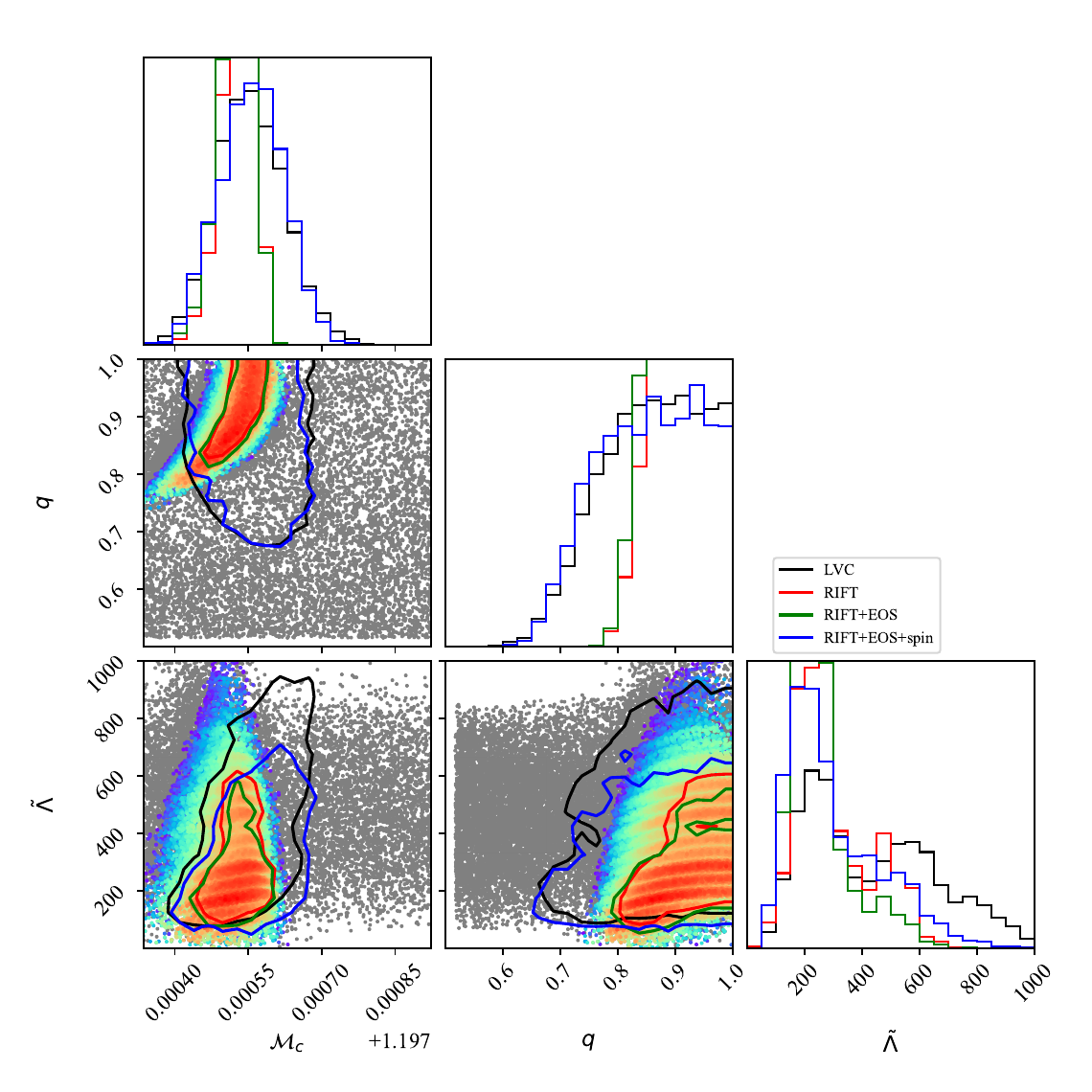}
\caption{\label{fig:170817_prototype}\textbf{Prototype analyses of GW170817}:
Corner plot with 90\% credible intervals  of the RIFT analysis. 
The color scale shows the likelihood range with red points indicating the highest likelihoods and blue points the
lowest, drawn from analyses assuming exactly zero spin.
The grey points portray the likelihoods below the cut-off value of dynamic likelihood range $\Delta \ln \mathcal{L} \leq 15$.
The panels show marginal distributions of the (detector-frame) chirp mass $\mathcal{M}_c$, the mass ratio $q$, and the effective tidal deformability $\tilde{\Lambda}$.
\textit{Legend}:
LVC (black line) - source properties of GW170817 with low spin ($|\chi| \leq 0.05$) from the initial LIGO/Virgo analysis;
RIFT (red line) - RIFT analysis of GW170817 with no spin and uniform prior on $\Lambda_i$;
RIFT+EOS (green line) - RIFT analysis of GW170817 with no spin and the EOS index linked to the Legred et al. EOS table; 
RIFT+EOS+spin (blue line) - RIFT analysis of GW170817 with spin, the EOS index linked to the Legred et al. EOS table.
}
\end{figure}

In Figure \ref{fig:170817_prototype}, we revisit the analysis of GW170817, using settings comparable to a previously
published analysis \cite{LIGO-GW170817-bns} (data release: \cite{gravitational_wave_open_science_center_strain_2018}).
Most of the marginal distributions, credible intervals, and likelihoods are constructed with RIFT for this work.  As a benchmark, the black curves
(labeled LVC) show a previously-published analysis of GW170817 \cite{LIGO-GW170817-bns}, using similar settings and priors except for allowing both NS spins to
have arbitrary orientation and uniformly distributed spin magnitude.   
We also assume that both compact objects are neutron stars.
As seen in this figure, the source properties of GW170817 with low spin ($|\chi| \leq 0.05$) from the initial LIGO/Virgo analysis are: a
source-frame chirp mass 
$\mathcal{M}_c = 1.188^{+0.004}_{-0.002} M_\odot $, a mass ratio $q=0.7 - 1.0$,  and a tidal deformability $\tilde{\Lambda} \leq
800$.
For comparison, the blue curves show our analysis of the same event using RIFT,  under the assumption that 
their EOS are drawn uniformly from the Legred set of prior EOS realizations (labeled RIFT+EOS+spin).  We adopt priors
uniform in the component masses and uniform in $\chi_{i,z}$ over the allowed range.    
Given our assumption of
  low NS spin magnitudes, the transverse spins have little effect on the orbit or emitted radiation: orbital precession is small as the orbit is
  highly angular momentum dominated \cite{ACST,2019PhRvD.100b4046S}.  Our uniform prior for $\chi_{i,z}$ differs modestly from the
appropriate one-dimensional marginal prior equivalent to uniform spin magnitude (the ``z
prior'') \cite{gwastro-PENR-RIFT}, less strongly penalizing large spins with opposite signs and thus producing a
broader and more forgiving marginal prior for the net effective inspiral spin ($\chi_{\rm eff}$)
\cite{2021arXiv210409508C}.  Due to correlations in the likelihood between $q$ and $\chi_{\rm eff}$, our more forgiving
spin prior allows for more asymmetric mass ratios.   That said, despite numerous differences in the underlying priors (precession vs spin; EOS) and analysis (RIFT
versus \texttt{lalinference}), the two analyses broadly agree, save that the EOS-conditioned analysis favors lower tidal deformabilities.

\begin{figure*}
\includegraphics[width=0.48\textwidth]{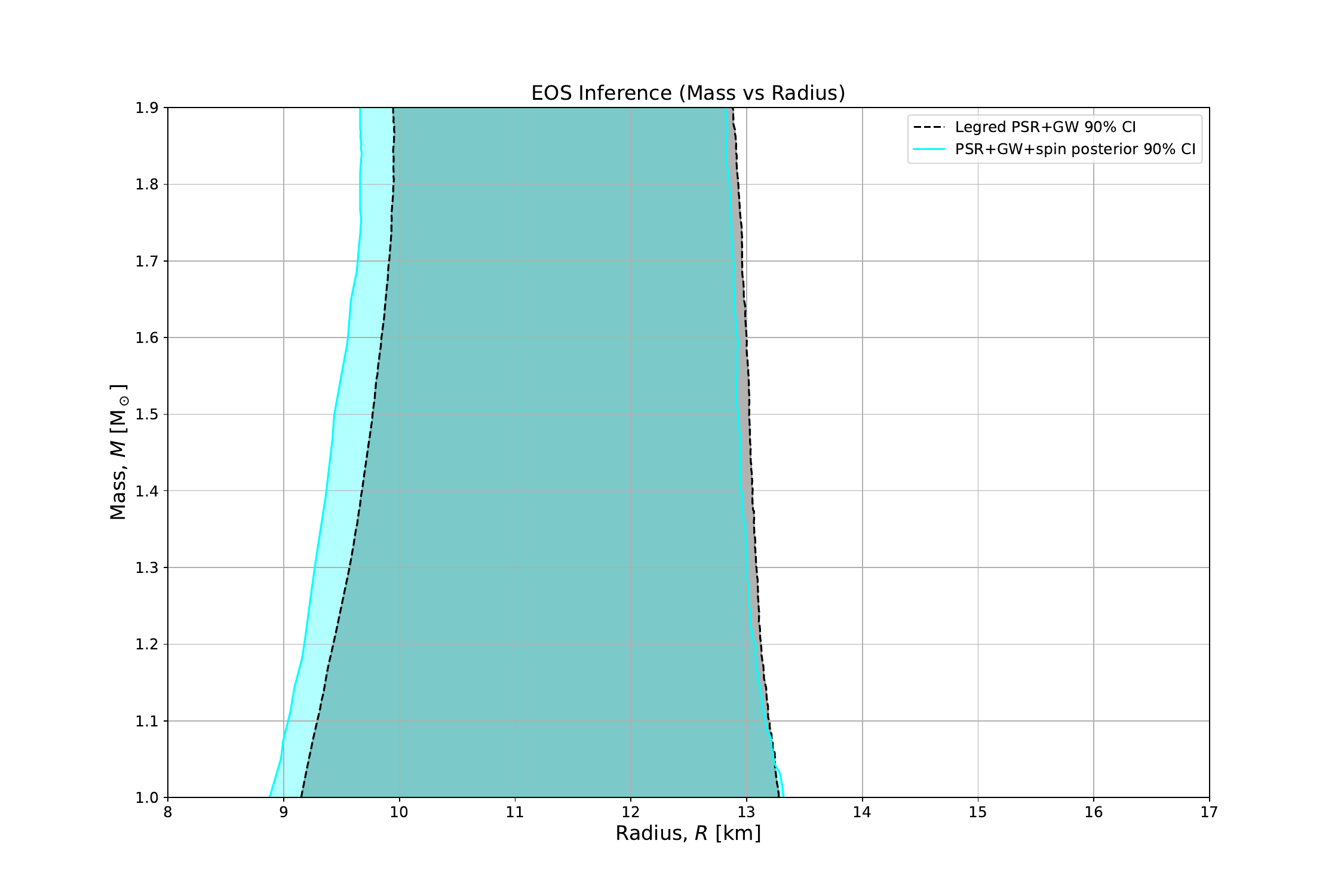}
\includegraphics[width=0.48\textwidth]{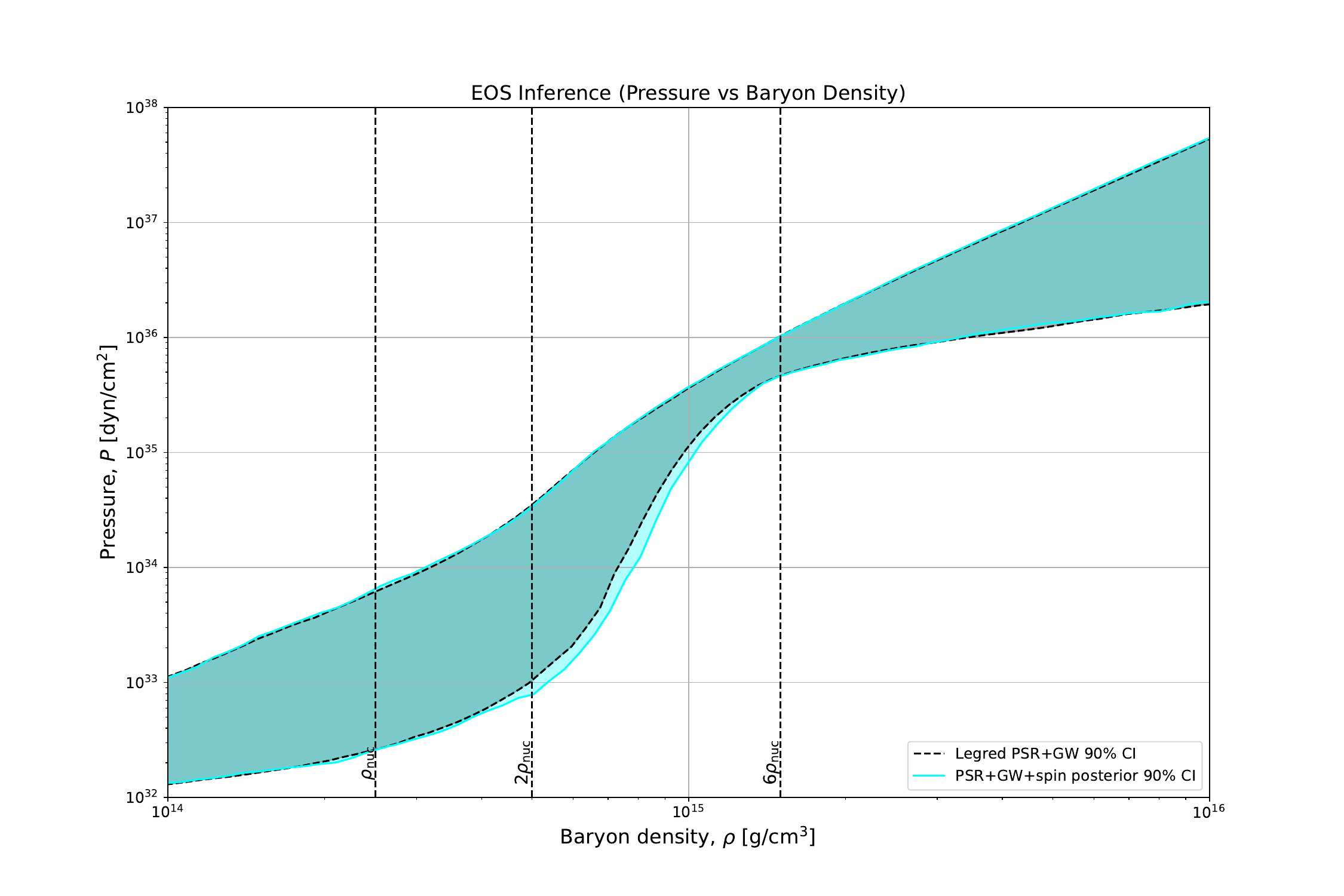}
\caption{\label{fig:170817_EOS_vs_Legred}\textbf{Inferred EOS from GW170817 - Validation}: Inferred EOS derived from our
  fiducial analysis of GW170817 seen in Figure \ref{fig:170817_prototype}, expressed as 90\%
  credible intervals in terms
  of pressure and density (left panel) and radius versus mass (right panel), rendered as radius verus mass, and labelled
  by ``PSR+GW+spin posterior''.  For
  comparison, this figure also shows the results of the Legred analysis (dashed black lines) based on the same tabular EOS
  family, combining pulsar and gravitational wave observations.  
}
\end{figure*}

Our model-based EOS analysis directly provides a posterior distribution over EOS realizations.  Figure \ref{fig:170817_EOS_vs_Legred} 
shows the inferred distribution of EOS in pressure versus density and radius versus mass, expressed as 90\% quantiles at
each fixed $\rho$ and $M$, respectively.    This figure also shows the results of comparable analysis based on an identical
initial tabular EOS family, derived using data products provided by Legred derived by comparing the same tabular EOS family to GW170817.
This agreement is striking because 
our analysis exploits slightly different information than the Legred analysis, which was conditioned on all GW
observations including GW190425, not just GW170817 as performed in this work.

\begin{figure*}
    \includegraphics[width=0.49\textwidth]{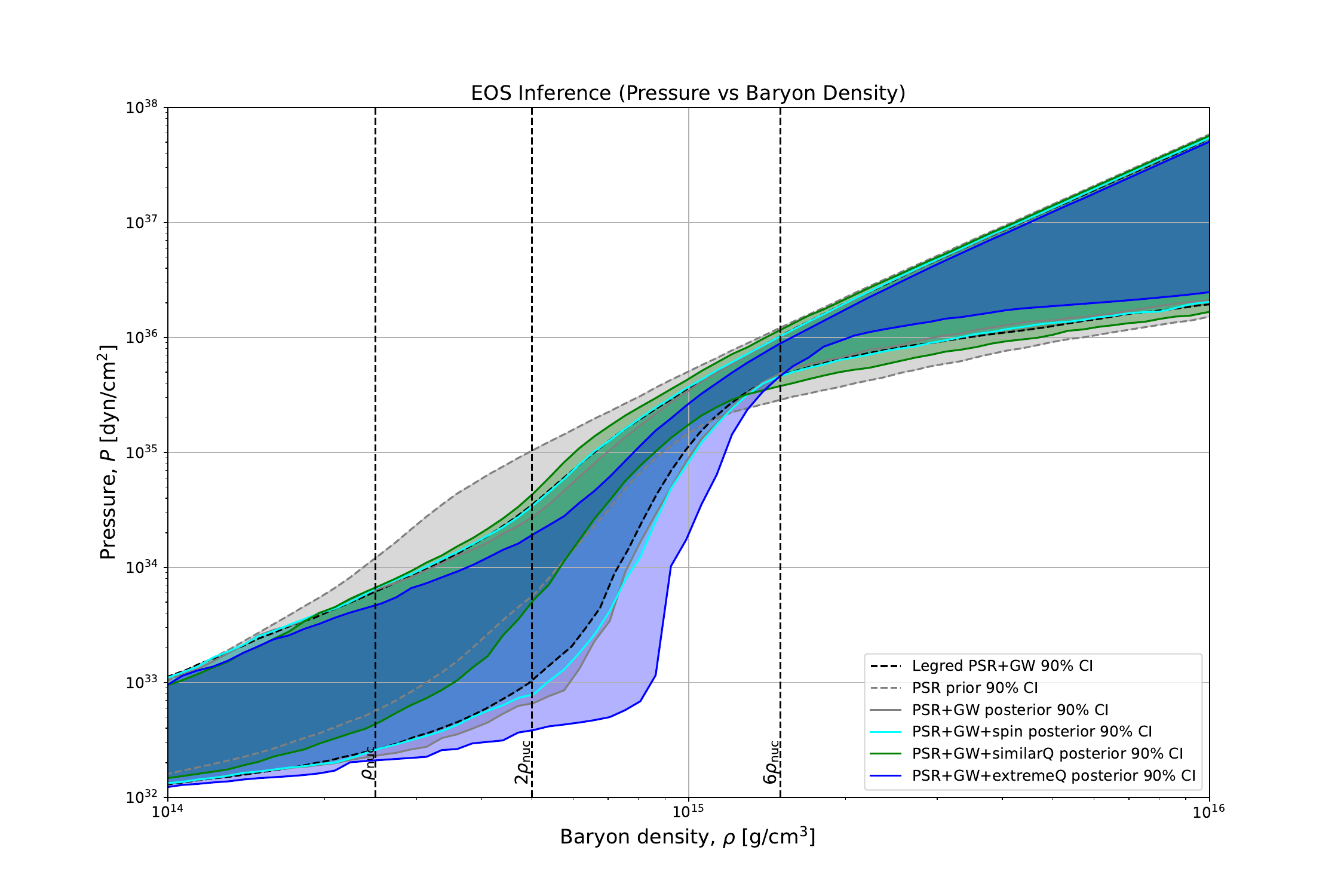}
    \includegraphics[width=0.49\textwidth]{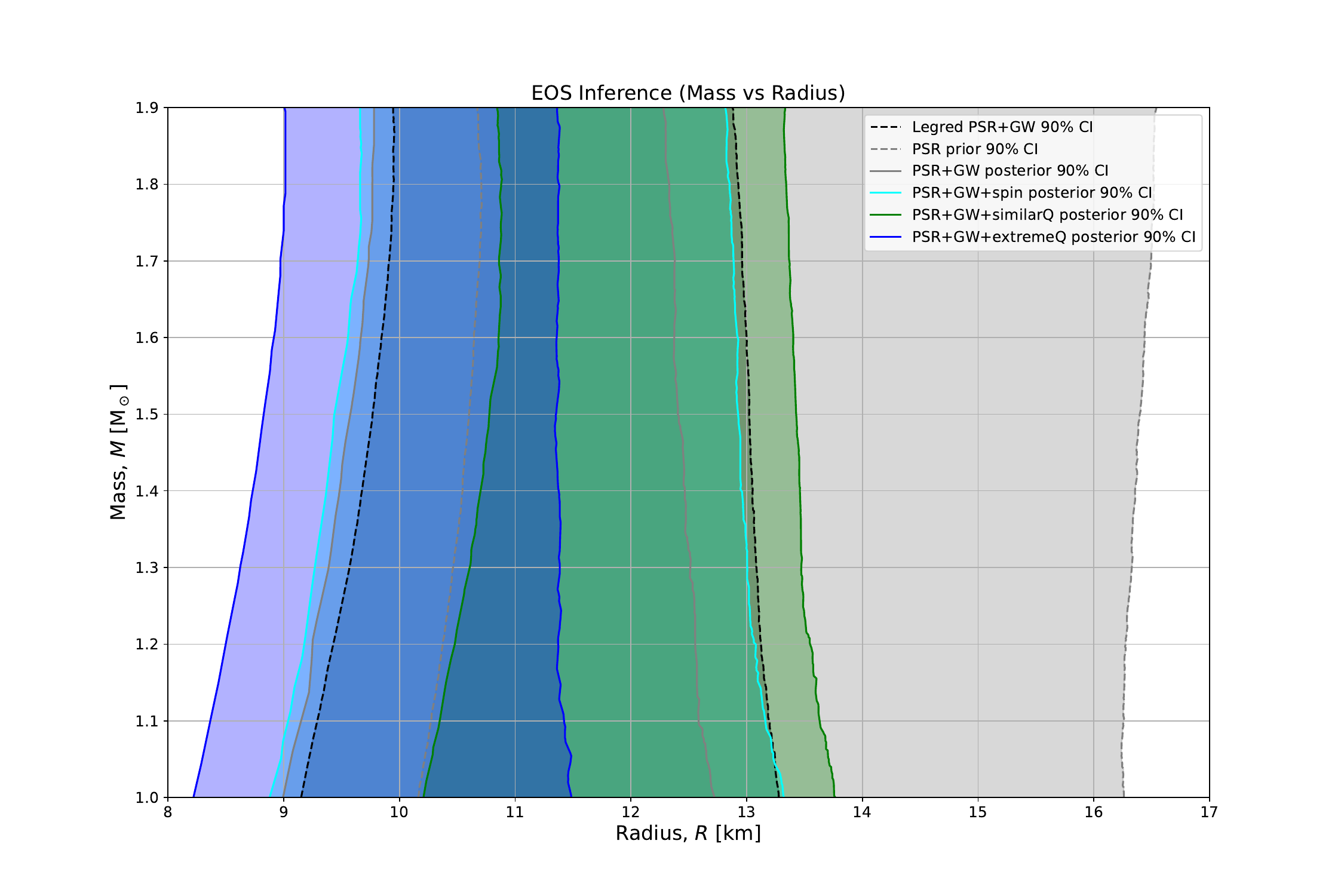}
\caption{\label{fig:170817_EOS}\textbf{Inferred EOS from GW170817}: Inferred EOS,
  showing both the prior and posterior distributions
\emph{Left panel}: 90\% credible interval for pressure at each fixed baryon density $\rho$.
\emph{Right panel}: 90\% credible interval for radius at each fixed mass, rendered as radius versus mass.
\emph{Legend}: 
Legred PSR+GW (black dashed line) - posterior distribution of the nuclear equation of state conditioned on pulsar and GW observations (Legred et al.);
PSR prior (grey dashed line) - posterior distribution of the nuclear equation of state conditioned on pulsar observations, used as a prior for RIFT analysis of GW170817;
PSR+GW (grey solid line) - posterior distribution of the nuclear equation of state conditioned on pulsar and GW observations with no spin assumption;
PSR+GW+spin (cyan solid line) - posterior distribution of the nuclear equation of state conditioned on pulsar and GW observations with spin;
PSR+GW+similarQ (green solid line) - posterior distribution of the nuclear equation of state conditioned on pulsar and GW observations with similar mass ratio;
PSR+GW+extremeQ (blue solid line) - posterior distribution of the nuclear equation of state conditioned on pulsar and GW observations with extreme mass ratio.
}
\end{figure*}

As previously demonstrated in many studies, adding  GW information significantly informs the EOS, both as represented in
pressure versus density or radius versus mass.  To illustrate the importance of GW information relative to the  broad
underlying EOS prior, Figure \ref{fig:170817_EOS} re-expresses our result and the corresponding prior 90\% credible  as 
 `PSR+GW+spin posterior' and  `PSR prior' .

\subsection{Isolating the impact of the EOS prior}
Having demonstrated our work reproduces the Legred et al EOS inference results, we perform a narrowly-focused
investigation to directly assess the impact of the EOS prior, compared to a conventional unmodeled phenomenological
approach where $\Lambda_i$ are drawn uniformly from a prior distribution.  To eliminate systematics associated with the
use of different inference codes and analysis settings, we perform all analyses with RIFT using identical settings and
priors.  Furthermore, to  reduce the complexity of our analysis, we have performed these analyses assuming both neutron
stars have exactly zero spin.   

Figure  \ref{fig:170817_prototype}  shows our results for source parameter inference.  The solid red credible intervals
and distributions (and point color
scale)  show our fiducial zero-spin inference for GW170817, allowing $\Lambda_i$ to be drawn uniformly between
$[0,5000]$.
By contrast, the green contours show inference informed by uniformly drawing from the Legred EOS prior draws.  By a
remarkable coincidence, the two posteriors are surprisingly similar versus $\tilde{\Lambda}$, suggesting the  naive
phenomenological prior has a remarkably similar effect as a well-motivated nonparametric prior over the support of the posterior.
Indeed, the nonparametric posterior for $\tilde{\Lambda}$ in Figure \ref{fig:170817_prototype} is  similar to both the zero-spin
EOS-informed analysis (RIFT+EOS) and even the spinning EOS-informed RIFT analysis (RIFT+EOS+spin).   Unsurprisingly, the
EOS inferences derived from the zero-spin analysis (Figure  \ref{fig:170817_EOS}) are extremely similar to the EOS
inferences from the full analysis presented above. 

To summarize, our
model-based approach to the EOS leads to a tidal deformability distribution comparable to an unmodeled phenomenological
approach.

\subsection{Alternative astrophysical priors}
Following Kedia et al \cite{gwastro-ns-eos-Kedia2024}, we also repeat our analyses using two alternative mass ratio and spin priors for
GW170817, with the same assumption that both components are neutron stars.  On the one hand, to
focus on high mass ratios only, we will perform some analyses with  a mass ratio distribution $p(q)$ that is consistent with a
uniform distribution of component masses but excludes $m_2/m_1 > 0.6$.  [At GW170817's source-frame chirp mass $\mc = 1.186 M_\odot $, this constraint
  requires more extreme choices for source-frame primary and secondary mass:  $m_2 < 1.062M_\odot$ and $m_1>1.77
  M_\odot$.]  
These extreme mass ratios are more easily supported by the data if NS can sometimes have large spins, so 
we  consider the nonprecessing equivalent of an
isotropic uniform spin magnitude prior which extends to $0.4$.  
Conversely, when requiring GW170817's mass ratio
prior distribution to be constrained to the range expected for GW-detected BNS sources derived from the galactic population, we require
 $m_2/m_1>0.9$   \cite{2020ApJ...900L..41A} and, to strongly disfavor extreme spin-orbit misalignment,
we simultaneously require $\chi_{i,z}>0$.    Operationally, all of these astrophysical prior choices can be implemented as  alternative
options for  the upper and lower
bounds adopted for $q,\chi_{i,z}$,  when analyzing GW170817 with a nonprecessing GW model.

\begin{figure*}
\includegraphics[width=0.9\textwidth]{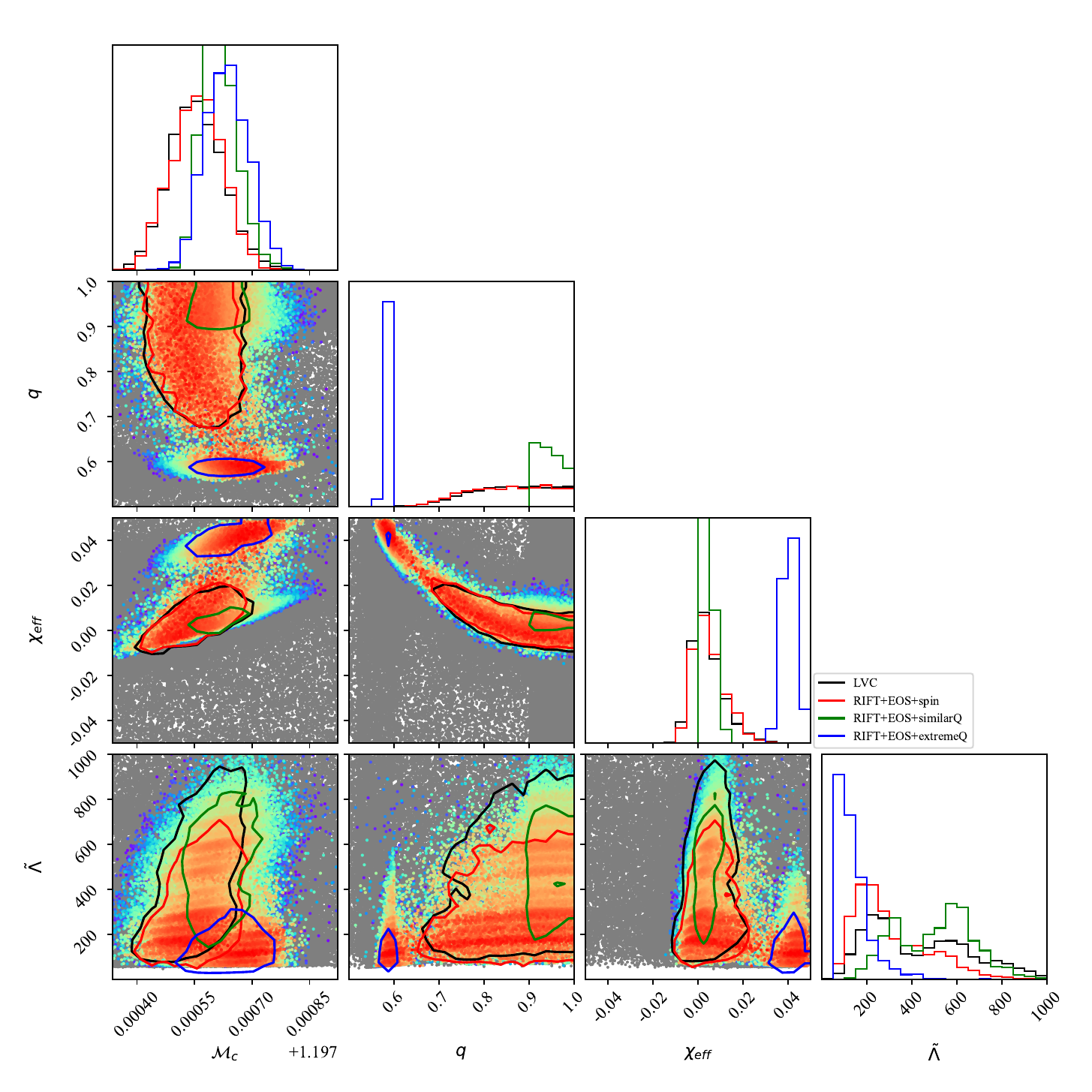}
\caption{\label{fig:170817_spin}\textbf{Analysis of GW170817 focused on different spin priors}:
Corner plot with 90\% credible intervals  of the RIFT analysis.
The color scale shows the likelihood range with red points indicating the highest likelihoods and blue points the lowest.
The grey points portray the likelihoods below the cut-off value of dynamic likelihood range $\Delta \ln \mathcal{L} \leq 15$.
The panels show marginal distributions of the (detector-frame) chirp mass $\mathcal{M}_c$, the mass ratio $q$, the effective spin value $\chi_eff$, and the effective tidal deformability $\tilde{\Lambda}$.
\emph{Legend}:
LVC (black line) - source properties of GW170817 with low spin ($|\chi| \leq 0.05$) from the initial LIGO/Virgo analysis;
RIFT+EOS+spin (red line) - RIFT analysis of GW170817 with spin, the EOS index linked to the Legred et al. EOS table;
RIFT+EOS+similarQ (green line) - RIFT analysis of GW170817 with similar mass ratio, and the EOS index linked to the Legred et al. EOS table;
RIFT+EOS+extremeQ (blue line) - RIFT analysis of GW170817 with extreme mass ratio, and the EOS index linked to the Legred et al. EOS table.
}
\end{figure*}

Figure \ref{fig:170817_spin} shows the results of our inferences for GW170817.  As previously, the solid black curve
shows the fiducial LIGO/Virgo analysis adopting phenomenological EOS priors.  In this figure,  the solid red curve shows
the RIFT+EOS+spin analysis previously presented in Figure \ref{fig:170817_prototype}.  The remaining two curves show the
two new analyses informed by our alternative spin priors: the green curve shows inferences obtained by assuming $q>0.9$
and $\chi_{i,z}>0$, while the blue curve requires $q<0.6$.  Both of these analyses are performed self-consistently with
a uniform prior over the Legred tabular EOS family. As in Kedia et al \cite{gwastro-ns-eos-Kedia2024}, the bottom right panel of this corner plot illustrates the
impact of these different astrophysical prior choices: the former (green) analysis favors large $\tilde{\Lambda}$, while
the latter (blue) favors small $\tilde{\Lambda}$.

Since we have performed self-consistent simultaneous EOS  and parameter inference, we can directly propagate the
impact of these different astrophysical priors into strikingly different conclusions about the EOS.  Figure
\ref{fig:170817_EOS} shows inferences (in green and blue, respectively) about the EOS, expressed in terms of pressure
versus density and mass versus radius.    As required, the two assumptions have opposite effects, pushing the EOS to a
notably larger or smaller radius.

Allowing for astrophysical prior uncertainty substantially broadens the range of possibilities for the radius $R_{1.4}$
of a $1.4 M_\odot$ neutron star.  For our fiducial PSR/GW/EOS-only analysis, we would nominally favor a 90\% credible
interval between $9.5-12.7\,\unit{km}$. 
When we add the spin to our analysis, the 90\% credible interval for $R_{1.4}$ is $9.2 - 13.0\,\unit{km}$.
The alternative astrophysical prior choices lead to credible intervals shifted towards the lower and upper ends of the $R_{1.4}$ range.
The extreme mass ratio analysis leads to a 90\% credible interval of $8.5 - 11.5 \,\unit{km}$, while the similar mass ratio analysis leads to a 90\% credible interval of $10.5 - 13.5 \,\unit{km}$.
Taking the union of all our 90\% credible intervals would suggest a fiducial NS radius $R_{1.4}$ between $8.5-13.5\,\unit{km}$,
derived using GW information and EOS prior information alone.

Our conclusions about the fiducial NS radius $R_{1.4}$ arrive at a significantly broader systematic interval  than the  Kedia et al approach for two key reasons.  On the one hand, our analysis
only uses GW information, not NICER information about NS radii.  On the other hand, Kedia's model-based analysis of
these events makes strong assumptions about the low-density EOS.  Both of these differences favor a narrower result in
Kedia et al's approach.

\section{Conclusions}
\label{sec:conclusions}

We have demonstrated a method to allow the RIFT parameter-inference engine to simultaneously compare (tabulated draws
from) arbitrary EOS model families to individual GW events, performing simultaneous source and parameter inference.   To
validate our approach, we replicated previous analyses of GW170817 and its nuclear EOS,  using tabulated draws from a
previous-published nonparametric family provided by Legred et al \cite{2021PhRvD.104f3003L}.   To demonstrate the utility of our
method, following Kedia et al \cite{gwastro-ns-eos-Kedia2024} we have further performed joint parameter and EOS inference for GW170817
under different astrophysical assumptions.  Even using the much more expressive gaussian process EOS family provided by
Legred et al, we again replicate Kedia's key conclusion: that different astrophysical assumptions about GW170817 will
produce starkly different conclusions about the nuclear EOS.

This method dramatically increases the scope of RIFT-based EOS inference, previously limited to parametric models and
largely to single-event analysis.  For example, by iteratively folding the results from one tabular analysis into the
next, we can draw conclusions based on prior information from previous and other analyses, such as inferences including
NICER and other GW sources. 
These EOS-informed source parameter inferences have extremely modest overhead, requiring comparable resources to a
conventional RIFT EOS-agnostic analysis of the same event.
Moreover, because this analysis still provides EOS-agnostic likelihood evaluations $(\lambda_\alpha, \ln {\cal
  L}_{\rm marg}(\lambda_\alpha))$, any  analyses performed with
this approach produce results that can be  transparently and directly incorporated in downstream analyses  which adopt
other EOS models.   In particular, these analyses produce likelihood samples that can be directly incorporated in multi-event and
multimessenger EOS and population inference \cite{gwastro-ns-eos-Kedia2024,2020arXiv200101747W}, without additional computational cost.

Algorithmically speaking, our method combines two technical advantages.  On the one hand, by using RIFT interpolated marginal likelihoods, we
can perform  inference using the full, multivariate, and highly correlated GW posterior.  Our method can in particular
perform inference even in regions strikingly lower in likelihood than the maximum likelihood.   By contrast, other methods
that bootstrap from fiducial posterior samples must carefully calibrate their approximations against those samples (e.g., KDE, normalizing
flow, et cetera).  Even then, these methods likely cannot extrapolate safely to regions not well sampled in that
fiducial analysis.  In particular, these methods will likely have difficulty replicating our analysis of strong
astrophysical priors, which probe regions of much lower likelihood where few to no  fiducial samples will  be available.
On the other hand, by using tabular EOS inference, we can directly employ the results of any state-of-the-art
theoretical or empirical investigation that provides EOS posterior samples, including methods that calibrate to
low-density models (e.g., crust models and chiral effective field theory) and apply high-density perturbative QCD
constraints; see, e.g., \cite{2024PhRvD.109i4030K,2024arXiv240204172K} and references therein.
Finally and not least, the RIFT framework is exceptionally efficient, allowing end-to-end inference in at most a few
days, even using EOS inference and even employing costly waveform models. 
In particular, our approach does not need a training phase before performing inference: 
the full self-consistent analysis can be performed in a single run.

\acknowledgments
The authors thank Phillipe Landry for a constructive and helpful review of the manuscript.  
ROS acknowledges support from NSF Grant No. AST-1909534, NSF Grant No. PHY-2012057, and the
Simons Foundation. AY acknowledges  support from NSF Grant No. PHY-2012057. AV acknowledges support from  AST-1909534.
JR acknowledges support from NSF RUI-2110441 and AST-2219109.
We are grateful for
the computational resources provided by the LIGO Laboratory, supported via National Science Foundation Grants PHY-0757058 and PHY-0823459.
This material is based upon work supported by NSF’s LIGO Laboratory which is a major facility fully funded by the
National Science Foundation.

\appendix

\section{Dependence on mass ratio} 
\label{app:mass-ratio-dependence}
\begin{figure*}
  \includegraphics[width=0.49\textwidth]{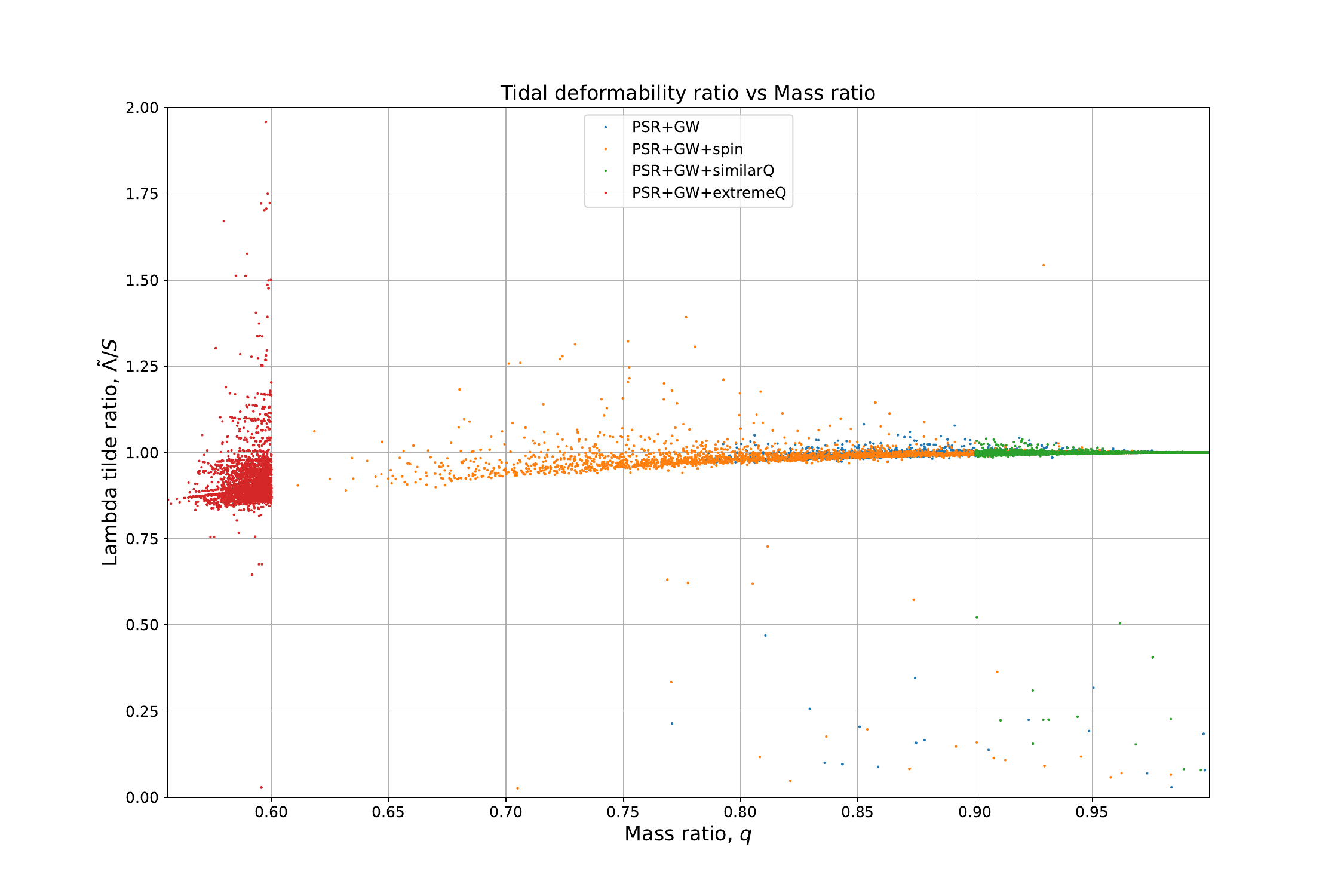}
  \includegraphics[width=0.49\textwidth]{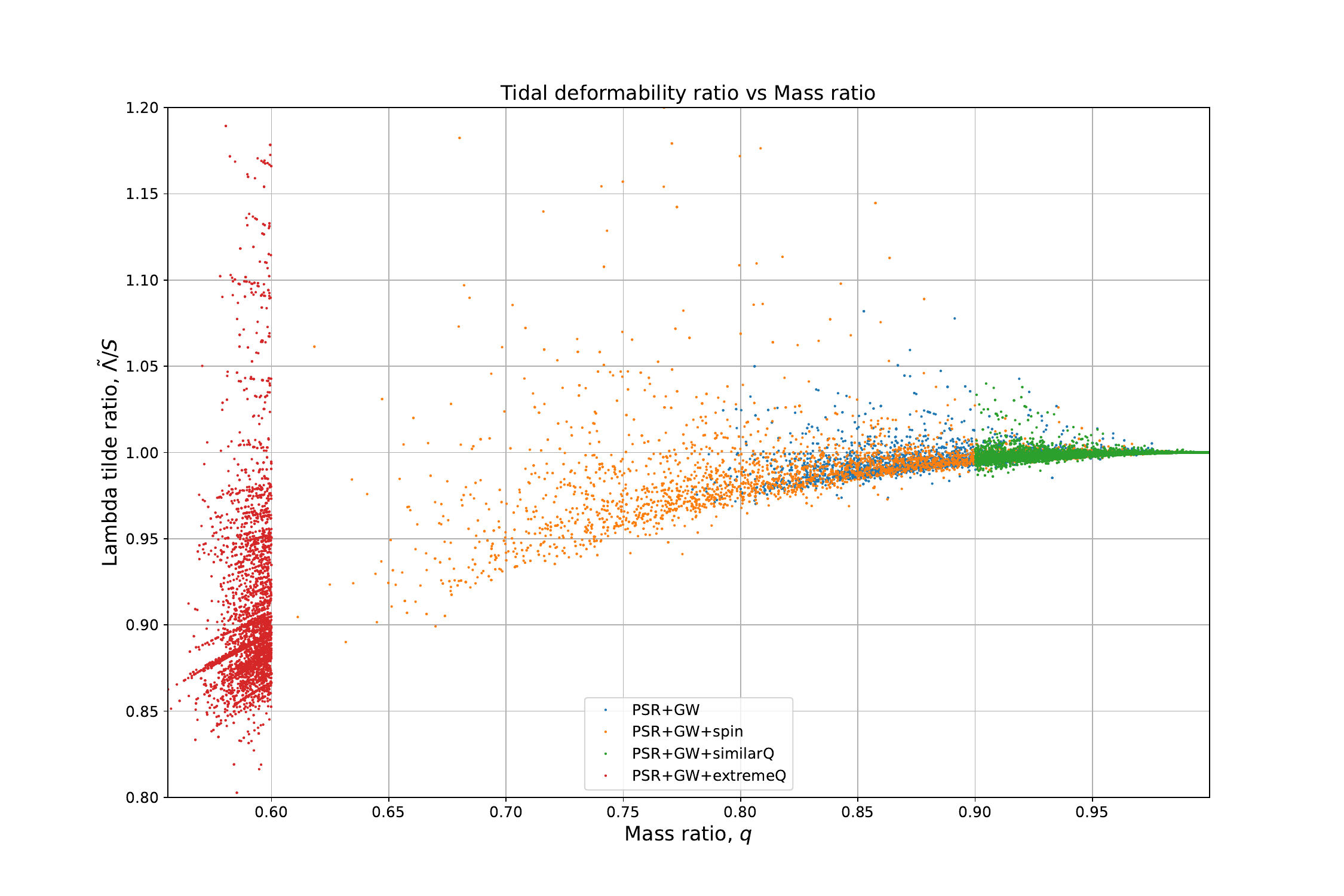}

  \caption{\label{fig:LambdaTilderatio_vs_q}\textbf{Tidal deformability ratio ($\tilde{\Lambda}$ to the ordering statistic $S$) vs mass ratio $q$}:
    The ratio of the tidal deformability $\tilde{\Lambda}$ to the ordering statistic $S$ as a function of mass ratio $q$.
    \emph{Left panel}: The ratio of $\tilde{\Lambda}$ to $S$ including the full range of EOS models.
    \emph{Right panel}: The ratio of $\tilde{\Lambda}$ to $S$ enlarged in the region where most of the data points are located.
    \emph{Legend}:
    PSR+GW (blue dots) - posterior samples conditioned on pulsar and GW observations with no spin assumption;
    PSR+GW+spin (orange dots) - posterior samples conditioned on pulsar and GW observations with spin;
    PSR+GW+similarQ (green dots) - posterior samples conditioned on pulsar and GW observations with similar mass ratio;
    PSR+GW+extremeQ (red dots) - posterior samples conditioned on pulsar and GW observations with extreme mass ratio.}
\end{figure*}

In the text, following previous work \cite{2018PhRvD..98f3020Z} we assume $\tilde{\Lambda}$ does not vary appreciabily
with mass ratio.  While this assumption allows us to employ only a one-dimensional ordering statistic assuming
$m_1=m_2$, at a high mass ratio this assumption is less valid.  To assess the utility of our assumption, we revisited the
results of our phenomenological and EOS-informed inference, using the posterior draws $m_i$ to identify candidate mass
configurations to investigate.    In each case,
for every EOS in our prior model from Legred et al, we compared $\tilde{\Lambda}$ calculated using our prior to the ordering statistic $S$
computed using our simplifying approximation.   Figure \ref{fig:LambdaTilderatio_vs_q} shows our result.  Even allowing
for the full prior range from Legred's Gaussian-process model family, our approach is usually within better than 5\% when
$q> 0.8$ and often within 10\% when $q>0.7$, consistent with prior work \cite{2018PhRvD..98f3020Z}.  Our extremely broad
approach does identify some outliers as one can see on the left panel of Fig. \ref{fig:LambdaTilderatio_vs_q}, but
their number is relatively small and they are mostly caused by the very extreme EOS models.
Such EOS models appear in the posterior distribution very rarely and largely overestimate the ordering statistic $S$ values 
(corresponding to the $\tilde{\Lambda}$ in the EOS models). 
On the other hand, extreme mass ratios result in lower $S$ values with similar $\mathcal{M}_c$ values, so there are more outliers with the higher ratio of $\tilde{\Lambda}$ to $S$.
Appendix \ref{app:R14_Lambda14} and Fig. \ref{fig:R14_Lambda14} show the posterior distributions for $R_{1.4}$ and $\Lambda_{1.4}$, 
which help to understand the effects these features have on the final results.

At more extreme mass ratios, the bias and dispersion of the $S,\tilde{\Lambda}$ correlation are both larger, introducing
additional systematic uncertainty into inferences derived for extreme-mass-ratio binaries. Because of the large
exponents relating $\tilde{\Lambda}$ and $R_{1.4}$, even an $O(10\%)$ bias here can produce changes of order a kilometer
or more to the final inferred NS radius. 
For instance, $R_{1.4}$ values for PSR+GW and PSR+GW+spin results are $10.85$ km and $11.15$ km, respectively, 
while the extreme mass results in $R_{1.4}$ values of $9.84$ km. 
For this reason, high-precision tabular EOS
inference using even observations from this regime will require going beyond our simplifying initial one-dimensional
approximation.

\section{Posterior distributions for $R_{1.4}$ and $\Lambda_{1.4}$}
\label{app:R14_Lambda14}

\begin{figure*}
  \includegraphics[width=0.49\textwidth]{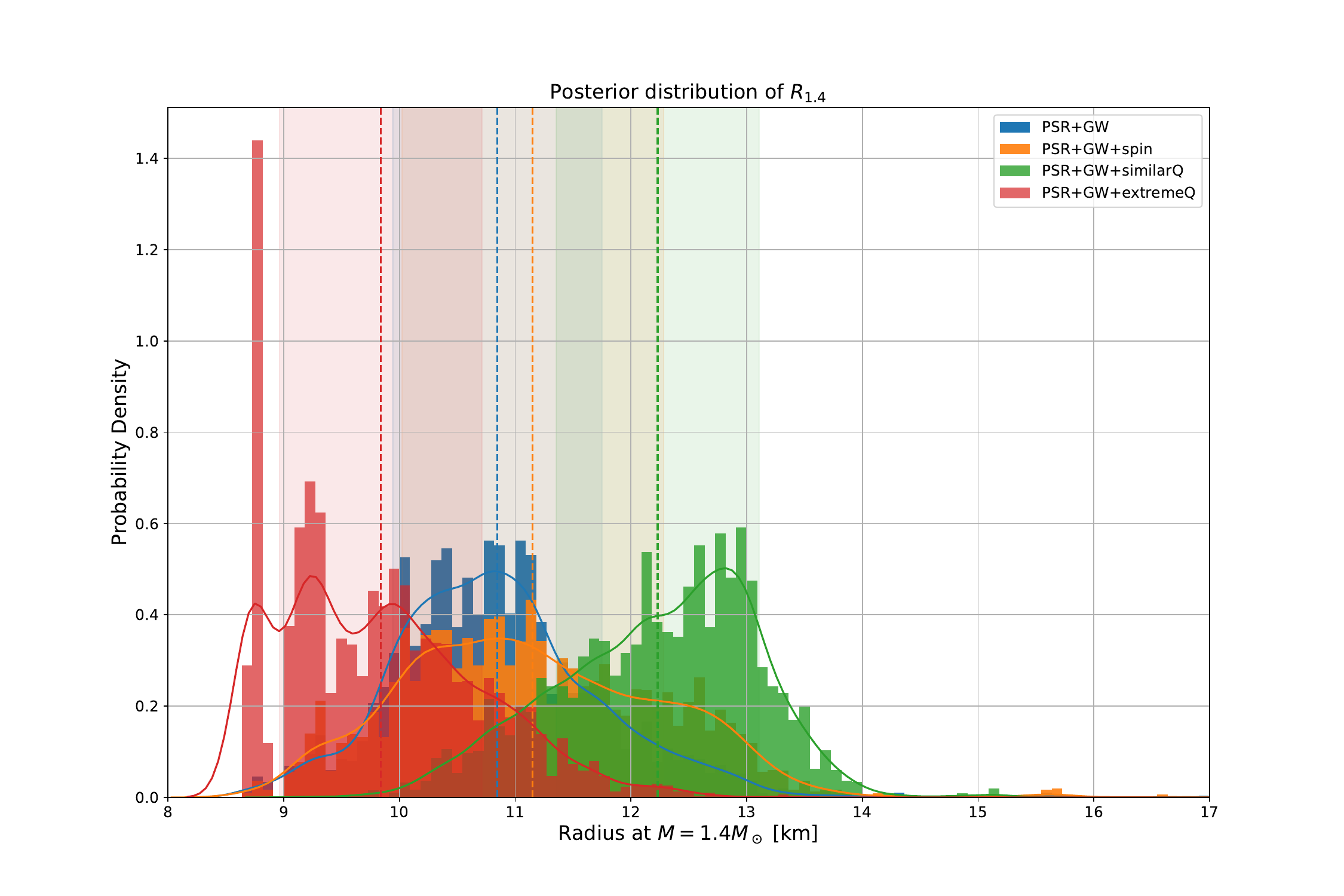}
  \includegraphics[width=0.49\textwidth]{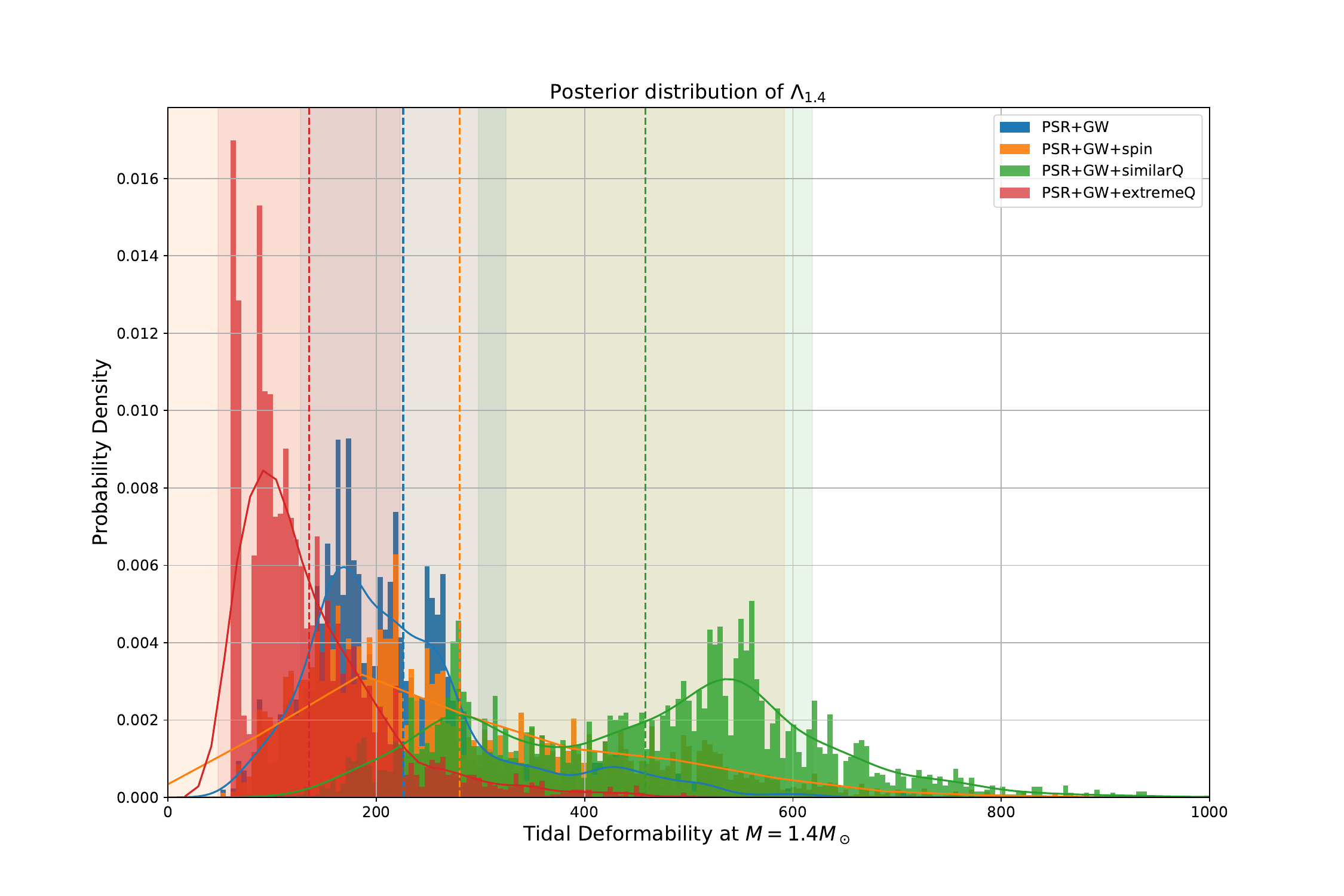}

  \caption{\label{fig:R14_Lambda14}\textbf{Posterior distributions for $R_{1.4}$ and $\Lambda_{1.4}$}:
    Posterior distributions for the radius $R$ and tidal deformability $\Lambda$ of a $1.4 M_\odot$ neutron star. 
    Dashed lines show the mean values and the shaded regions show the standard deviations of the posterior distributions.
    Please note that the $\Lambda_{1.4}$ values here correspond to the ordering statistic $S$ used inside the analysis.
    \emph{Left panel}: Posterior distribution for the radius $R_{1.4}$ of a $1.4 M_\odot$ neutron star.
    \emph{Right panel}: Posterior distribution for the tidal deformability $\Lambda_{1.4}$ of a $1.4 M_\odot$ neutron star.
    \emph{Legend}:
    PSR+GW (blue) - posterior distribution conditioned on pulsar and GW observations with no spin assumption;
    PSR+GW+spin (orange) - posterior distribution conditioned on pulsar and GW observations with spin;
    PSR+GW+similarQ (green) - posterior distribution conditioned on pulsar and GW observations with similar mass ratio;
    PSR+GW+extremeQ (red) - posterior distribution conditioned on pulsar and GW observations with extreme mass ratio.}
\end{figure*}

In this section, we present the posterior distributions for the radius $R$ and tidal deformability $\Lambda$ of a $1.4 M_\odot$ neutron star. 
As one can see in Fig. \ref{fig:R14_Lambda14}, the PSR+GW, PSR+GW+spin, and PSR+GW+extremeQ analyses generally agree with each other. 
Mean values of PSR+GW and PSR+GW+spin are consistent with each other, while the PSR+GW+extremeQ analysis results in slightly lower values: 
$R_{1.4} = 10.85$ km and $\Lambda_{1.4} = 236$ for PSR+GW, 
$R_{1.4} = 11.15$ km and $\Lambda_{1.4} = 292$ for PSR+GW+spin, and
$R_{1.4} = 9.84$ km and $\Lambda_{1.4} = 136$ for PSR+GW+extremeQ.
The similar mass ratio analysis PSR+GW+similarQ results in significantly higher values - $R_{1.4} = 12.23$ km and $\Lambda_{1.4} = 465$.
The reason for such difference is that the similar mass ratio analysis covers only a small region of mass ratio $q > 0.9$, 
where the tidal deformability $\Lambda$ tends to be higher. 
PSR+GW and PSR+GW+spin analyses, on the other hand, cover a wider range of mass ratios. 
Thus, their results are averaged over a wider range of $\Lambda$ values, 
where lower mass ratios take up a larger portion of the posterior distribution.

As discussed in Appendix \ref{app:mass-ratio-dependence}, the approximation of the mass ratio $q$ as a single value $q=1$ might cause problems in the high mass ratio region.
In the left panel of Fig. \ref{fig:R14_Lambda14}, one can see that the PSR+GW+extremeQ analysis has three peaks in the posterior distribution of $R_{1.4}$. 
This shows the uncertainty discussed in the previous section, which results from the broad dispersion of the $\tilde{\Lambda}$ to $S$ ratio in the high mass ratio region shown in Fig. \ref{fig:LambdaTilderatio_vs_q}.
For comparison, the PSR+GW, PSR+GW+spin, and PSR+GW+similarQ analyses do not show such abrupt changes in the posterior distribution of $R_{1.4}$. 
Also, in the left panel of the figure one can see that the PSR+GW+spin analysis has two peaks that are consistent with the PSR+GW analysis and the PSR+GW+similarQ analysis.

\bibliography{paperexport,references,LIGO-publications,gw-astronomy-mergers,gw-astronomy-mergers-approximations, gw-astronomy-mergers-ns-nuclearphysics}
\end{document}